\documentclass[10pt]{IEEEtran}

\usepackage{amsmath,amssymb}
\usepackage{graphicx}
\usepackage{verbatim}
\usepackage{multirow}
\usepackage{enumerate}
\usepackage{cite}

\title{Reduced-Complexity Decoder of Long Reed-Solomon Codes Based on Composite
  Cyclotomic Fourier Transforms}

\author{Xuebin Wu and Zhiyuan Yan}

\begin{document}
\maketitle

\begin{abstract}
  Long Reed-Solomon (RS) codes are desirable for digital communication
  and storage systems due to their improved error performance, but the
  high computational complexity of their decoders is a key obstacle to
  their adoption in practice. As discrete Fourier transforms (DFTs)
  can evaluate a polynomial at multiple points, efficient DFT
  algorithms are promising in reducing the computational complexities
  of syndrome based decoders for long RS codes.
  In this paper, we first propose partial composite cyclotomic Fourier
  transforms (CCFTs) and then devise syndrome based decoders for long
  RS codes over large finite fields based on partial CCFTs. The new
  decoders based on partial CCFTs achieve a significant saving of
  computational complexities for long RS codes. Since partial CCFTs
  have modular and regular structures, the new decoders are suitable
  for hardware implementations. To further verify and demonstrate the
  advantages of partial CCFTs, we implement in hardware the syndrome
  computation block for a $(2720, 2550)$ shortened RS code over
  GF$(2^{12})$. In comparison to previous results based on Horner's
  rule, our hardware implementation not only has a smaller gate count,
  but also achieves much higher throughputs.


\end{abstract}
\section{Introduction}

Since syndrome-based hard-decision decoders of Reed-Solomon (RS) codes
\cite{BlahutECC} have quadratic complexities in their code lengths, RS
codes of short and medium lengths have widespread applications in
modern digital communication and storage systems. To meet ever higher
demand on error performance, long RS codes (or shortened RS codes
\cite{Buerner2004,Jeon2010}) over large finite fields have been
considered in theoretical studies. For example, RS codes with
thousands of symbols over GF$(2^{12})$ are considered in optical
communication systems \cite{Buerner2004} and magnetic recording
systems \cite{Han2005,IDEMAWhitePaper} to achieve low bit error
rates. One of the key obstacles to the adoption of such long RS codes
in practice is high complexity caused by their extreme code lengths as
well as the large sizes of their underlying fields.

Fast algorithms for discrete Fourier transforms (DFTs) over finite
fields are promising techniques to overcome this obstacle. This is
because all steps except the key equation solver in syndrome-based
hard-decision RS decoders \cite{BlahutECC} --- syndrome computation,
Chien search, and error magnitude evaluation --- are polynomial
evaluations. Hence, they can be formulated as DFTs over finite fields.

Recently, cyclotomic fast Fourier transforms (CFFTs) over finite
fields have been used to reduce the complexities of RS decoders
\cite{Costa2004,Chen2009}.  CFFTs proposed in
\cite{Trifonov2003,Costa2004,Fedorenko2006} have low multiplicative
complexities, but they have very high additive complexities.  By using
techniques such as the common subexpression elimination (CSE)
algorithm in \cite{Chen2009c}, the additive complexities of CFFTs can
be significantly reduced, leading to small overall computational
complexities for DFTs with lengths up to $1024$ \cite{Chen2009c}.  By
treating syndrome computation, Chien search, and error magnitude
evaluation as partial CFFTs or dual partial CFFTs, the overall
computational complexities of these steps can be significantly reduced
for short and medium RS codes \cite{Costa2004,Chen2009}.
Unfortunately, this approach will not be feasible for long DFTs and
hence long RS codes. This is because the CSE algorithm itself has a
prohibitively high computational complexity when applied to long
DFTs. Without the CSE algorithm, the overall computational
complexities of CFFTs will be higher than other approaches due to
their additive complexities.

In this paper, we devise reduced-complexity decoders for long RS codes
based on composite cyclotomic Fourier transforms (CCFTs)
\cite{Wu2011a}.  CCFTs first decompose long DFTs with composite
lengths into short sub-DFTs via the prime-factor algorithm
\cite{Good1958} or the Cooley-Tukey algorithm \cite{Cooley1965}, and
then implement the sub-DFTs with CFFTs.  We remark that CFFTs are
special cases of CCFTs corresponding to trivial decompositions.  The
decomposition leads to significantly reduced additive complexities at
the expense of multiplicative complexities, resulting in lower overall
computational complexities than CFFTs for moderate to long DFTs in
practice \cite{Wu2011a}.  Furthermore, the decomposition also endows
CCFTs with modular structures, which are suitable for hardware
implementations.

The main contributions of this paper are  as follows:
\begin{itemize}
\item We first propose partial CCFTs and then apply them to implement
  syndrome computation, Chien search, and error magnitude evaluation
  of RS decoders. Partial CCFTs not only inherit the two advantages
  (lower additive complexities and modular structures) of full CCFTs,
  their two-tier structure is also suitable for the implementation of
  decoders for shortened RS codes. For instance, for DFTs in shortened
  RS codes, certain time-domain elements are zeros and certain
  frequency-domain components are not needed. For partial CFFT, either
  property can lead to multiplicative complexity reduction but not
  both at the same time.  The two-tier structure of CCFT, however,
  enables us take advantage of both properties simultaneously to
  reduce the multiplicative complexity. Consequently, our results show
  that partial CCFTs leads to a significant saving of computational
  complexities for long RS codes.
\item To further verify and demonstrate the advantages of partial
  CCFTs, we implement in hardware the syndrome computation block for a
  $(2720, 2550)$ shortened RS code over GF$(2^{12})$. In comparison to
  previous results based on Horner's rule, our hardware implementation
  not only has a smaller gate count, but also achieves much higher
  throughputs.
\end{itemize}

The rest of this paper is organized as follows. We review CFFTs and
CCFTs in Sec.~\ref{sec:background}.
Sec.~\ref{sec:result} first proposes partial CCFTs and then presents
RS decoders based CCFTs. The hardware implementation results are
provided in Sec.~\ref{sec:VLSI}. Finally, our paper concludes in
Sec.~\ref{sec:conclusion}.


\section{Background}
\label{sec:background}
\subsection{CFFTs and CCFTs over Finite Fields}
Assuming that $\alpha \in \mbox{GF}(2^m)$ is an element of order $n$,
the DFT of an $n$-dimensional vector $\mathbf{f}=(f_0, f_1, \cdots,
f_{n-1})^T$ over GF$(2^m)$ is given by $\mathbf{F}=(f(\alpha^0),
f(\alpha^1), \cdots, f(\alpha^{n-1}))^T$, where $f(x) =
\sum_{i=0}^{n-1}f_ix^i$. That is, DFTs can be viewed as polynomial
evaluations.  The vector $\mathbf{f}$ is said to be in the time domain
and $\mathbf{F}$ in the frequency domain. Direct CFFTs (DCFFTs)
\cite{Trifonov2003} formulate the DFTs as $\mathbf{F}=\mathbf{ALf}'$,
where $\mathbf{A}$ is an $n\times n$ binary matrix, $\mathbf{L}$ a
block diagonal matrix with each block cyclic, and $\mathbf{f}'$ a
permutation of $\mathbf{f}$. Since the multiplication between a cyclic
matrix and a vector can be done by efficient bilinear algorithm of
cyclic convolution, CFFTs can be computed by
$\mathbf{F}=\mathbf{A}\mathbf{Q}(\mathbf{c}\cdot
\mathbf{P}\mathbf{f})$, where $\mathbf{Q}$ and $\mathbf{P}$ are binary
matrices, $\mathbf{c}$ is a pre-computed vector, and $\cdot$ denotes
an entry-wise multiplication between two vectors. Two variants of
DCFFTs, referred to as inverse CFFTs (ICFFTs) \cite{Costa2004} and
symmetric CFFTs (SCFFTs) \cite{Fedorenko2006}, respectively, compute
the DFTs by $\mathbf{F}=\mathbf{L}^{-1}\mathbf{A}^{-1}\mathbf{f}'$ and
$\mathbf{F}=\mathbf{L}^T\mathbf{A}^T\mathbf{f}'$, respectively. Since
it has been shown that ICFFTs and SCFFTs are equivalent
\cite{Chen2009c}, without loss of generalization we consider only
DCFFTs and SCFFTs in this paper.

The composite cyclotomic Fourier transform in \cite{Wu2011a} can
further reduce the overall computational complexity by decomposing the
long DFTs into short sub-DFTs via the prime-factor algorithm
\cite{Good1958} or the Cooley-Tukey algorithm \cite{Cooley1965}.  The
decompositions of the DFTs reduce the additive computational
complexity directly. Moreover, because of the short length of the
sub-DFTs, sophisticated tools such as the CSE algorithm in
\cite{Chen2009c}, can be readily used to reduce the additive
complexities of CCFTs.  CCFTs also have a modular structure, which is
desirable in hardware implementation. The sub-DFTs can be used as
sub-modules, which can be reused to save chip area or parallelized to
increase the throughput.

\subsection{Reed-Solomon Decoders based on CFFTs}
Henceforth in this paper, we focus on cyclic Reed-Solomon (RS) codes,
which can be decoded by syndrome-based decoders considered herein
\cite{BlahutECC}. For an $(n, k)$ cyclic RS code over GF$(2^m)$ with
$n|2^m-1$ and $n-k=2t$, it can correct up to $t$ errors or $2t$
erasures. An $(n',k')$ shortened RS code can be viewed as a sub-code
of an $(n, k)$ RS code where the symbols at the position $i\ge n'$ are
always zero. For a received vector $\mathbf{r}=(r_0, r_1, \cdots,
r_{n-1})^T$, the syndrome-based errors-only (errors-and-erasures,
respectively) decoder of RS codes in the time domain consists of the
following three steps \cite{BlahutECC}:
\begin{enumerate}
\item Compute the $2t$ syndromes $s_j=\sum_{i=0}^{n-1}r_i\alpha^{ij}$ for
  $0\le j \le 2t-1$, where $\alpha$ is an $n$-th primitive element.
\item Compute the error (errata) locator polynomial $\Lambda(x)$ and
  error (errata) evaluator polynomial $\Omega(x)$ by the Berlekamp-Massey
  algorithm (BMA) or the extended Euclidean algorithm.
\item Find the error (errata) positions by the Chien search. That is,
  the error positions are obtained by finding the root of
  $\Lambda(x)$. Find the error (errata) value by Forney's formula,
  which evaluates $\Omega(x)$ and $\Lambda'(x)$ (formal derivative of
  $\Lambda(x)$) at the error (errata) positions.
\end{enumerate}

Since evaluating a polynomial at multiple points can be implemented as
a DFT, DFTs can be used to reduce the computational complexity of
steps 1 and 3.  When DFTs are used to implement syndrome computation
in the RS decoder, only $2t$ frequency-domain elements are
needed. Hence, the unnecessary rows and columns of the matrices in
DCFFTs or SCFFTs can be removed to reduce both multiplicative and
additive complexities, resulting in partial DCFFTs and partial
SCFFTs. Similarly, when DFTs are used to evaluate the error (errata)
locator and evaluator polynomials, many time-domain elements are
zeroes due to the limited degrees of both polynomials. Again the
unnecessary rows and columns of the matrices in DCFFTs and SCFFTs can
be removed, leading to dual partial DCFFTs and dual partial SCFFTs.
Since a shortened RS code is essentially a RS code with zero symbols,
these zero symbols are treated as zero time-domain elements. When DFTs
are used to implement syndrome computation, the Chien search, and
Forney's formula, these DFTs are partial in both time and frequency
domains.

Although the complexity of the Berlekamp-Massey algorithm is important
to efficient RS decoders, the implementation of the Berlekamp-Massey
algorithm is not considered henceforth in this paper, since the
computational complexity of the Berlekamp-Massey algorithm cannot be
reduced by DFTs.

\section{RS Decoders Based on Partial Composite Cyclotomic Fourier Transforms}
\label{sec:result}
In this section, we first propose partial CCFTs and then devise
syndrome-based time-domain RS decoder based on our partial CCFTs. The
complexities of our RS decoder are compared with previous works in the
literature.

\begin{figure}[!htb]
  \centering
  \includegraphics[width=5.5cm]{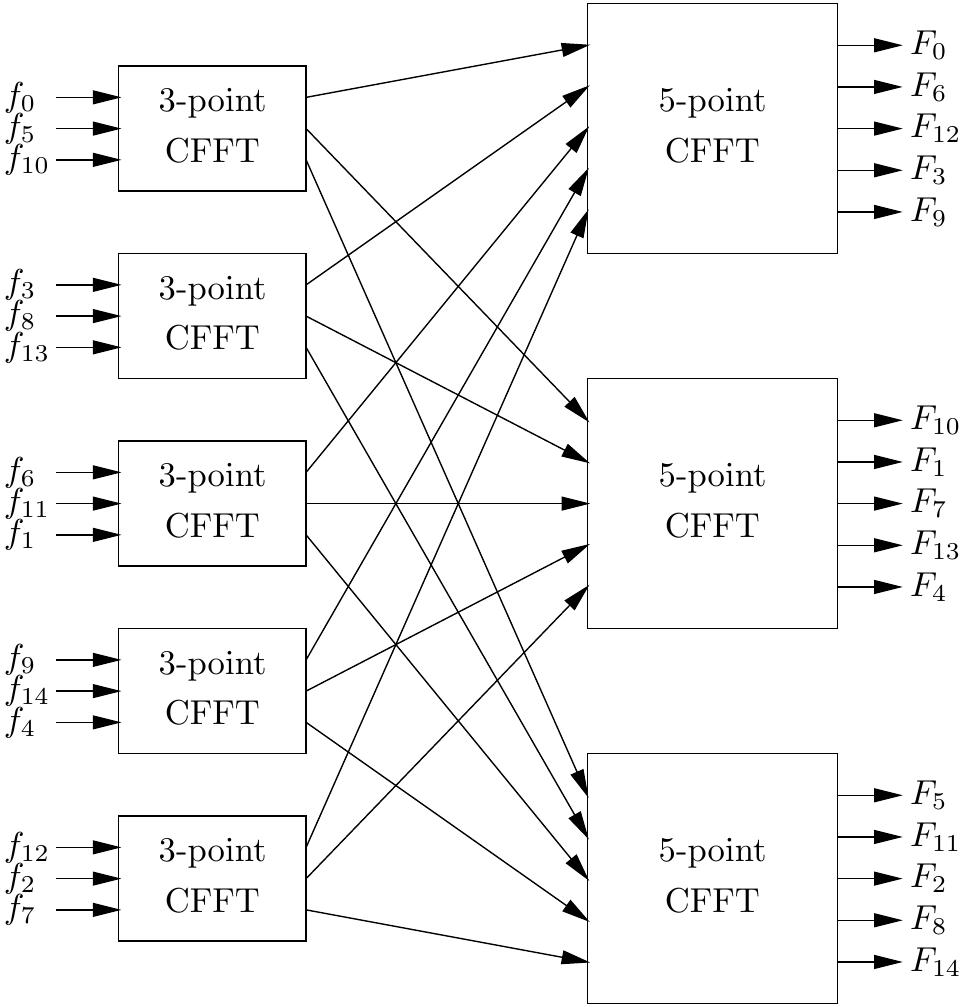}
  \caption{The regular and modular structure of our $15$-point CCFT based on a $3 \times 5$ decomposition. }
  \label{fig:CCFT3x5}
\end{figure}

\subsection{Partial Composite Cyclotomic Fourier Transforms}
\label{sec:partialCCFT}

When $N=N_1N_2$, with the prime-factor algorithm \cite{Good1958} or
the Cooley-Tukey algorithm \cite{Cooley1965}, an $N$-point CCFT can be
carried out in a two-tier structure. The first tier performs $N_2$
$N_1$-point CFFTs and the second performs $N_1$ $N_2$-point
CFFTs. When the greatest common divisor of $N_1$ and $N_2$ is greater
than one, twiddle factors are needed. When $N_1$ and $N_2$ are
co-prime to each other, no twiddle factor is required. When $N_1$ or
$N_2$ is composite, $N_1$- or $N_2$-point DFTs can be further
decomposed, leading to multi-tier structure.  Fig.~\ref{fig:CCFT3x5}
shows the two-tier structure of a $3 \times 5$ CCFT, where the first
tier consists of five 3-point CFFTs and the second tier three 5-point
CFFTs.  This regular and modular structure is suitable for hardware
implementations, since it is much easier to apply architectural
techniques such as folding and pipelining to this regular and modular
structure, leading to efficient hardware implementations.

When some frequency-domain components are not needed or some of the
time-domain elements are always zeroes, the corresponding rows and
columns of matrices in the sub-CFFTs can be removed, resulting in
partial CCFTs. As shown in \cite{Wu2011a}, CCFTs have lower
computational complexities than CFFTs in evaluating long DFTs, and
hence we expect that partial CCFTs have advantages in reducing the
computational complexities of decoders for long RS codes.

We remark that if we decompose an $N$-point DFT as $1\times N$, the
corresponding partial CCFT will reduce to partial SCFFT, and if we
decompose the DFT as $N\times 1$, the corresponding partial CCFT will
reduce to partial DCFFT. Therefore, our partial CCFTs include partial
DCFFTs and partial SCFFTs as special cases. In this sense, DFT
decomposition provides another degree of freedom to reduce the
computational complexities of DFTs. In the following, we focus on the
computational complexities of partial CCFTs with non-trivial
decompositions, i.e., decompositions other than $1\times N$ and
$N\times 1$.

We discuss the complexity of partial CCFTs can be reduced based on
partial time or frequency domain elements, and compare partial CCFTs
with partial CFFTs. Assuming a two-tier structure for simplicity,
there are three possible scenarios: \begin{enumerate}
\item \textbf{When limited frequency domain elements are needed}.\\
  For RS codes, when DFTs are used to compute the syndromes of a
  received vector, only the first $2t$ frequency-domain components are
  needed. The results in \cite{Chen2009} show that the multiplicative
  complexity of a partial SCFFT is reduced greatly, but because the
  matrix $\mathbf{A}$ is not sparse, it is hard to reduce the
  multiplicative complexity of a partial DCFFT. Even though partial
  DCFFTs have smaller additive complexities than partial SCFFTs, they
  have higher overall computational complexities. For partial CCFTs,
  the multiplicative complexity of the second tier can be directly
  reduced due to the unnecessary frequency-domain components. However,
  since computing even one frequency-domain component of an
  $N_2$-point vector requires all of the time-domain elements, the
  outputs of the DFTs in the first tier may only have unnecessary
  frequency-domain components in some rare cases, e.g., the number of
  the DFTs in the second tier is more than that of the necessary
  frequency-domain components, and hence the complexity of the DFTs in
  the first tier cannot be reduced in most cases. Thus, the complexity
  reduction of partial CCFTs is not as great as partial CFFTs.

\item \textbf{When some time domain elements are zero}.\\
  For RS codes, when DFTs are used to reduce the computational
  complexities of Chien search and error evaluation, only a few time
  domain components are non-zero, and hence partial DCFFTs can reduce
  the multiplicative complexities greatly and have lower overall
  complexities. For partial CCFTs, the multiplicative complexity of
  the first tier can be directly reduced due to the zero time domain
  components, while the complexity of the second tier cannot be easily
  reduced unless in rare cases.

\item \textbf{When limited frequency domain elements are needed and
    some time domain elements are zero}.\\ For shortened RS codes,
  only part of the time-domain elements are nonzero and only part of
  the frequency-domain components are needed. Neither partial DCFFTs
  nor partial SCFFTs can take full advantage of both properties
  simultaneously.  In contrast, the two-tier structure of partial
  CCFTs is advantageous. Due to the two-tier structure of CCFT, we can
  use DCFFTs in the first tier and SCFFTs in the second tier to reduce
  the multiplicative complexities as well as the overall complexities.
\end{enumerate}

\textbf{Example 1}: Consider a $(15, 11)$ RS code over GF$(2^4)$ with
a generator polynomial $\prod_{i=0}^{3}(x-\alpha^i)$, where $\alpha$
is a root of the primitive polynomial $x^4+x+1$. This code can correct
up to two errors or four erasures, and hence we need to compute the
first four frequency-domain components in the DFT of a received
codeword as the syndrome. We can decompose the 15-point DFT as
$3\times 5$ CCFT by the prime-factor algorithm as shown in
Fig.~\ref{fig:CCFT3x5}. The 3-point SCFFT in the first tier is given
by
$$
\begin{bmatrix}
  F^{(3)}_0 \\ F^{(3)}_1 \\ F^{(3)}_2 
\end{bmatrix}
=
\begin{bmatrix}
  1 & 0 & 0 & 0 \\
  0 & 1 & 1 & 0 \\
  0 & 1 & 0 & 1 
\end{bmatrix}
\left(
  \begin{bmatrix}
    \alpha^0 \\ \alpha^5 \\ \alpha^0 \\ \alpha^0 
  \end{bmatrix} \cdot
  \begin{bmatrix}
    1 & 1 & 1 \\
    0 & 1 & 1 \\
    1 & 0 & 1 \\
    1 & 1 & 0 \\
  \end{bmatrix}
  \begin{bmatrix}
    f^{(3)}_0 \\ f^{(3)}_1 \\ f^{(3)}_2
  \end{bmatrix}
\right),
$$
and the 5-point DCFFT in the second tier is given by
$$
\begin{bmatrix}
  F^{(5)}_0 \\F^{(5)}_1 \\ F^{(5)}_2 \\ F^{(5)}_3 \\ F^{(5)}_4  
\end{bmatrix}
=
\begin{bmatrix}
  1 0 0 0 0 0 0 0 0 1 \\
  1 1 1 1 1 0 0 0 0 0 \\
  1 1 1 0 0 0 0 1 1 1 \\
  1 1 1 0 0 1 1 1 1 0 \\
  1 1 1 1 1 1 1 0 0 0 \\
\end{bmatrix}
\left(
  \begin{bmatrix}
    \alpha^0 \\ \alpha^{10} \\ \alpha^{9} \\ \alpha^{8} \\ \alpha^0 \\
    \alpha^{10} \\ \alpha^{0} \\ \alpha^{4} \\ \alpha^{0} \\
    \alpha^{0} 
  \end{bmatrix}
  \cdot
  \begin{bmatrix}
    1 0 0 0 0 \\
    0 1 1 0 0 \\
    0 1 1 1 1 \\
    0 1 0 0 1 \\
    0 0 1 0 0 \\
    0 1 1 1 1 \\
    0 0 1 1 0 \\
    0 0 1 1 0 \\
    0 0 0 0 1 \\
    0 1 1 1 1 \\    
  \end{bmatrix}
  \begin{bmatrix}
    f^{(5)}_0 \\f^{(5)}_1 \\ f^{(5)}_2 \\ f^{(5)}_3 \\ f^{(5)}_4      
  \end{bmatrix}
\right).
$$
Since we need to compute the first four frequency components, from
Fig.~\ref{fig:CCFT3x5} we need the first and the fourth output from
the first 5-point CCFT module, the second output from the second one,
and the third output from the third one. Then the 5-point DFT modules
can be simplified by removing the unnecessary computations
accordingly. For example, when we simplify the first 5-point DFT
module, We can remove the second, third, and fifth rows in
$\mathbf{AQ}$, resulting in the fourth and fifth column containing
only zero. Then the corresponding rows in $\mathbf{c}$ and
$\mathbf{P}$ can be removed, thus reducing the additive and
multiplicative complexities. This is a similar reduction procedure
with the partial CFFT. However, the DFT modules in the first tier
cannot be simplified because all the outputs of these modules are
required for the computation in the second tier.

\textbf{Example 2}: Now let us consider a $(15, 13)$ RS code which can
correct one error or up to two erasures. Only the first two
frequency-domain components are needed and we still decompose the
15-point DFT by the prime-factor algorithm as $3\times 5$ CCFT. From
Fig~\ref{fig:CCFT3x5}, no output from the third 5-point DFT module is
needed and hence it can be removed. Therefore, the last output from
each 3-point DFT modules in the first tier is not needed, and hence
they can be simplified by removing unnecessary computations
accordingly. Only in this kind of cases, i.e., the number of the
required frequency-domain components is less than the number of DFTs
in the second tier, the computational complexity of the first tier can
be reduced.

\textbf{Example 3}: Consider a $(10, 6)$ RS code shortened from a
$(15, 11)$ code. In the syndrome computation step, we still need four
frequency components, which implies the 5-point DFTs in the second
tier can be simplified in the same way with Example 1. Moreover, as
the input $f_{10}$, $f_{11}$, $\cdots$, $f_{14}$ are zero, the
$3$-point DFT modules in the first tier connecting to these inputs can
be accordingly simplified.

These examples are relatively small, and they do not have smaller
complexities than the corresponding partial CFFTs. However, we can
expect that the partial CCFT will have smaller computational
complexity as the length of RS code increase.

\begin{table*}[!t]
  \centering
  \caption{Complexity comparison of the syndrome
    computation for errors-and-erasures RS decoders.}
  \label{tab:syncomp}
  \addtolength{\tabcolsep}{-0.5pt}
  \begin{tabular}{|c|c|cccc|ccc|ccc|cc|}
    \hline
    \multirow{2}{*}{Field} & \multirow{2}{*}{code} & \multicolumn{4}{c|}{Partial CCFT} &
    \multicolumn{3}{c|}{Partial SCFFT  \cite{Chen2009}} &
    \multicolumn{3}{c|}{Prime-factor \cite{Truong2006}} &
    \multicolumn{2}{c|}{Horner's rule \cite{BlahutECC}}\\
    \cline{3-14}
    & & $n_1\times n_2$ & Mult. & Add. & Total &  Mult. & Add. & Total &  Mult. & Add. &
    Total & Mult. & Add. \\
    \hline
    GF$(2^{8})$ & (255, 223) & $3\times 85$ & 252 & 2652 & 6432 & 149
    &  3970  & \textbf{6205} & 852 & 1804 & 14584 & 7874 & 8128\\
    GF$(2^{9})$ & (511, 447) & $7\times 73$ & 873 & 7268 & \textbf{22109} & 345
    & 16471 & 22336 & 5265 & 7309 & 35496 & 32130 & 32640 \\
    GF$(2^{10})$ & (1023, 895) & $31\times 33$ & 2868 & 18569 & \textbf{73061}
    & 824 & 60471 & 76397 & 6785 & 15775 & 144690 & 129794 & 130816 \\ \hline \hline
    GF$(2^{12})$ & (2720, 2550) & $63\times 65$ & 7565 & 63869 &
    \textbf{237864} & 1467  & 1244779 & 1278520 & -- & -- & -- &
    459511 & 462230\\
    GF$(2^{12})$ & (3073, 2731) & $63\times 65$ & 9268 & 82684 &
    \textbf{295848}  & 2782  & 2760210 & 2824196 & --  & --  & -- &
    1047552 & 1050624\\
    \hline
  \end{tabular}
\end{table*}

\subsection{Syndrome Computation}
\label{sec:syndrome}

For an $(n,k)$ RS code, the syndromes of a received vector
$\mathbf{r}=(r_0,r_1,\cdots,r_{n-1})^T$ are given by
$S_j=\sum_{i=0}^{n-1}r_i\alpha^{ij}$ for $0 \le j \le 2t-1$, which are
the first $2t$ frequency domain elements of the DFT of $\mathbf{r}$
and can be computed with our partial CCFT. For an $(n', k')$ RS codes
shortened from the $(n, k)$ codes, we can still use the $n$-point
partial CCFT to compute the syndrome, provided that the time-domain
elements of the CCFT input with indexes $i \ge n'$ are set to
zero. The partial CCFT can be then simplified correspondingly by
removing the unnecessary computations.

Due to their widespread applications, we select the $(255,223)$,
$(511, 447)$, and $(1023, 895)$ RS codes over GF$(2^8)$, GF$(2^9)$,
and GF$(2^{10})$, respectively, as examples to show computational
complexity reduction by partial CCFTs.  We also select two shortened
RS codes with parameters $(2720, 2550)$ \cite{Buerner2004} and $(3073,
2731)$ \cite{Jeon2010} over GF$(2^{12})$ to illustrate the advantage
of the two-tier structure.  

We compare the complexities of syndrome computation for the five RS
codes mentioned above based on partial CCFTs, partial SCFFT ,
prime-factor algorithm \cite{Truong2006}, and Horner's rule
\cite{BlahutECC} in Tab.~\ref{tab:syncomp}. For partial CCFTs, we have
tried all possible decompositions of the DFT lengths, and only the
non-trivial decompositions with the smallest computational
complexities are listed in Tab.~\ref{tab:syncomp}.  Note that due to
the extreme code length, the additive complexities of the syndrome
computation for the two shortened RS codes over GF$(2^{12})$ based on
partial CFFTs are not optimized with the CSE algorithm in
\cite{Chen2009c}. The total complexity in Tab.~\ref{tab:syncomp} is
defined to be a weighted sum of the additive and multiplicative
complexities. We assume that one multiplication has the same
complexity as $(2m-1)$ additions over the same field. This assumption
comes from both the hardware and software considerations
\cite{Chen2009c}. In Tab.~\ref{tab:syncomp}, the smallest total
complexities for all the codes are in boldface.

From Tab.~\ref{tab:syncomp}, we can see that both partial CCFTs and
partial SCFFTs have much smaller complexities than the Horner's rule,
which is used widely in practice. In GF$(2^8)$, partial CCFT have a
higher multiplicative complexity than partial SCFFT. However, due to
the reduced additive complexities, partial CCFTs have advantages in
smaller overall computational complexities in GF$(2^m)$ when $m= 9$ or
$10$, although the improvement is marginal in GF$(2^9)$ and
GF$(2^{10})$, roughly 1\% and 4\%, respectively. Due to the
sub-optimality of the CFFT and the efficiency of the CCFT for long
DFTs, the savings will be greater for larger fields. For the two
shortened RS codes over GF$(2^{12})$, the total complexities based on
partial CCFTs are only a \textbf{fraction} of those based on partial
CFFTs.

\begin{table*}[!t]
  \caption{Complexity comparison of combined
    Chien search and Forney's formula for errors-and-erasures RS decoders.}
  \label{tab:ChienForney}
  \centering
  \addtolength{\tabcolsep}{-2pt}
  \begin{tabular}{|c|c|c|ccccc|cccc|cccc|}
    \hline
    \multirow{2}{*}{Field} & \multicolumn{2}{c|}{\multirow{2}{*}{code}} & \multicolumn{5}{c|}{Partial CCFT} &
    \multicolumn{4}{c|}{Partial DCFFT} & \multicolumn{4}{c|}{Horner's
      Rule \cite{BlahutECC}}\\
    \cline{4-16}
    & \multicolumn{2}{c|}{}&$n_1\times n_2$ & Mult. & Add. & Div. &
    Total &  Mult. & Add. & Div. & Total & Mult. & Add. & Div. & Total\\
    \hline
    \multirow{4}{*}{GF$(2^{8})$} & \multirow{4}{*}{(255, 223)} & $\Omega(x)$ &
    $85\times 3$ & 252 & 2764 & 0 & 6544 & 149 & 3226 & 0 &
    \textbf{5461} & 992 & 992 & 0 & 15872\\
    \cline{3-16}
    & & $\Lambda_e(x)$ &  $85\times 3$ & 177 & 1845 & 0 & 4500 & 78 &
    1828 & 0&\textbf{2998}  & 4064 & 4080 & 0 & 65040\\
    \cline{3-16}
    & & $\Lambda_o(x)$ &  $85\times 3$ & 191 & 2230 & 0 & 5095 & 108 &
    3096 & 0 & \textbf{4716} & 4064 & 3825 & 0 & 64785\\
    \cline{3-16}
    & & Misc & & 0 & 255 & 32 &  & 0 & 255 & 32 & & 0 & 255 & 32 & \\
    \cline{3-16}
    & & Total & \multicolumn{9}{c|}{13175 + 32 divisions} &
    \multicolumn{4}{c|}{145952 + 32 divisions}\\
    \hline
    \multirow{4}{*}{GF$(2^{9})$} & \multirow{4}{*}{(511, 447)} & $\Omega(x)$ &
    $73\times 7$ & 834 & 6013 & 0 & 20191 & 345 & 12791 & 0 &
    \textbf{18656} & 4032 & 4032 & 0 & 72576 \\
    \cline{3-16}
    & & $\Lambda_e(x)$ &  $73\times 7$ & 658 & 4353 & 0 & 15539 & 177 &
    7802 & 0 & \textbf{10811} & 16320 & 16352 & 0 & 293792\\
    \cline{3-16}
    & & $\Lambda_o(x)$ &  $73\times 7$ & 678 & 4684 & 0 & \textbf{16210} & 248 &
    12533 & 0 & 16749 & 16320 & 15841 & 0 & 293281\\
    \cline{3-16}
    & & Misc & & 0 & 511 & 64 &  & 0 & 511 & 64 & & 0 & 511 & 64 &\\
    \cline{3-16}
    & & Total & \multicolumn{9}{c|}{46188 + 64 divisions } & \multicolumn{4}{c|}{660160 + 64 divisions}\\
    \hline
    \multirow{4}{*}{GF$(2^{10})$} & \multirow{4}{*}{(1023, 895)} &
    $\Omega(x)$ &  $33\times 31$ & 2687 & 16743 & 0 & \textbf{67796} & 824 &
    52557 & 0 & 68213 & 16256 & 16256 & 0 & 325120 \\
    \cline{3-16}
    & & $\Lambda_e(x)$ &  $33\times 31$ & 2295 & 14718 & 0 & 58323 & 430 &
    30294 & 0 & \textbf{38464} & 65408 & 65472 & 0 &  1308224 \\
    \cline{3-16}
    & & $\Lambda_o(x)$ &  $33\times 31$ & 2291 & 14523 & 0 &
    \textbf{58052} & 541 &
    51655 & 0 & 61934 & 65408 & 64449 & 0 & 1307201 \\
    \cline{3-16}
    & & Misc & & 0 & 1023 & 128 &  & 0 & 1023 & 128  & & 0 & 1023 & 128 &\\
    \cline{3-16}
    & & Total & \multicolumn{9}{c|}{165335 + 128 divisions} &
    \multicolumn{4}{c|}{2941568 + 128 divisions}\\
    \hline
    \multirow{4}{*}{GF$(2^{12})$} & \multirow{4}{*}{(2720, 2550)} &
    $\Omega(x)$ &  $65\times 63$ & 7807 & 65253 & 0 & \textbf{244814} & 1542 &
    1326289 & 0 & 1361755 &28730 & 28730 & 0 & 689520\\
    \cline{3-16}
    & & $\Lambda_e(x)$ &  $65\times 63$ & 6889 & 57631 & 0 & \textbf{216078} &
    787 & 691858 & 0 & 709959 & 231115 & 231200 & 0 &5546845\\
    \cline{3-16}
    & & $\Lambda_o(x)$ &  $65\times 63$ & 6897 & 57095 & 0 & \textbf{215726} &
    1082 &  1320622 & 0 & 1345508 & 231115 & 228480 & 0 & 5544125 \\
    \cline{3-16}
    & & Misc & & 0 & 2720 & 170 & & 0 & 2720 & 170 & & 0 & 2720 & 170 &\\
    \cline{3-16}
    & & Total & \multicolumn{9}{c|}{679338 + 170 divisions } &
    \multicolumn{4}{c|}{11780490+170 divisions}\\
    \hline
    \multirow{4}{*}{GF$(2^{12})$} & \multirow{4}{*}{(3073, 2731)} &
    $\Omega(x)$ &  $65\times 63$ & 9610 & 77852 & 0 & \textbf{298882} & 2908 &
    2760306 & 0 & 2827190 & 116622 & 116622 & 0 & 2798928\\
    \cline{3-16}
    & & $\Lambda_e(x)$ &  $65\times 63$ & 8033 & 66641 & 0 & \textbf{251400} &
    1550 & 1497544 & 0 & 1533194 & 525312 & 525483 & 0 & 12607659 \\
    \cline{3-16}
    & & $\Lambda_o(x)$ &  $65\times 63$ & 8018 & 65968 & 0 & \textbf{250382} &
    2041 & 2751557 & 0 & 2798500 & 525312 & 522410 & 0 & 12604586 \\
    \cline{3-16}
    & & Misc & & 0 & 3073 & 342 &  & 0 & 3073 & 342 & & 0 & 3073 & 342
    & \\
    \cline{3-16}
    & & Total & \multicolumn{9}{c|}{803737 + 342 divisions} &
    \multicolumn{4}{c|}{28014246 + 342 divisions}\\
    \hline
  \end{tabular}
\end{table*}

\subsection{Chien Search and Error Magnitude Evaluation}
\label{sec:ChienForney}

In RS decoders, the Chien search is used to determine the error (errata)
locations by finding the roots of the error (errata) locator polynomial
$\Lambda(x)$. It is implemented by evaluating $\Lambda(x)$ at all
points $\alpha^i$ in the finite fields GF$(2^m)$ with $0 \le i \le
2^m-2$, which can be done efficiently by fast DFT algorithms such as
partial CCFT in our paper. The input vector of the DFT only has at
most $2t+1$ nonzero elements. For shortened $(n', k')$ RS codes,
possible error (errata) locations must be less than $n'$. Therefore,
only the first $n'$ frequency-domain components are needed, and hence
partial CCFT can be simplified accordingly.

For the RS codes we study, Forney's formula \cite{BlahutECC} is
given by $Y_i = -\frac{\Omega(x)}{x\Lambda'(x)}\big|_{x=\alpha^{-j}},$
where $Y_i$ is the error (errata) magnitude at the $i$-th error
(errata) located at position $j$, and $\Lambda'(x)$ is the formal derivative
of $\Lambda(x)$. Although we evaluate $\Omega(x)$ and $\Lambda(x)$
only at the points corresponding to the error locations, the error
locations are variable from one received vector to another. Therefore,
we can evaluate $\Omega(x)$ and $\Lambda'(x)$ at all the points in the
finite field using partial CCFT, and then select the frequency-domain
components corresponding to the error locations.

Moreover, we can combine the computation of the Chien search and
Forney's formula by splitting the polynomial $\Lambda(x)$ into
$\Lambda_e(x)+\Lambda_o(x)$, where $\Lambda_e(x)$ and $\Lambda_o(x)$
are the sums of the terms in $\Lambda(x)$ with even and odd
degrees, respectively. It is easy to verify that in GF$(2^m)$,
$x\Lambda'(x)=\Lambda_o(x)$. Hence we can first evaluate the three
polynomials $\Omega(x)$, $\Lambda_e(x)$, and $\Lambda_o(x)$ at all
points in the finite field by partial CCFT, and then compute
$\Lambda(a)$ by $\Lambda_e(a)+\Lambda_o(a)$ for all $a \in
\mbox{GF}(2^m)$ with $n$ additional additions. The error locations are
the points where $\Lambda(x)=0$. With Forney's formula, the error
(errata) magnitudes can be computed with at most $t$ divisions ($2t$
divisions).

In Tab.~\ref{tab:ChienForney}, we compare the computational complexity
of combined Chien search and Forney's formula based on partial CCFTs
with non-trivial decompositions, partial DCFFTs, and Horner's rule for
the five RS codes and shortened RS codes discussed in
Sec.~\ref{sec:syndrome}.
The choices of partial CCFTs and CFFTs do not affect the number of
divisions. Similar to syndrome computation, the advantage of using
partial CCFTs (with non-trivial decompositions) instead of partial
CFFTs is rather limited for RS codes over GF$(2^m)$ when $m\leq
10$. However, the advantage of partial CCFTs is much greater in larger
fields. Again for the two shortened RS codes over GF$(2^{12})$, the
total complexities based on partial CCFTs are only a \textbf{fraction}
of those based on partial CFFTs.  Finally, since partial CFFTs are
special cases of partial CCFTs with trivial decomposition, we can
choose the most efficient algorithm to evaluate $\Omega(x)$,
$\Lambda_e(x)$, and $\Lambda_o(x)$, respectively. In
Tab.~\ref{tab:ChienForney}, the total complexity of combined Chien
search and Forney's formula based on partial CCFTs/CFFTs is also
provided.



\section{Hardware Implementations}
\label{sec:VLSI}
The additive and multiplicative complexities derived in
Sec.~\ref{sec:result} considers only the total number of the additions
and multiplications required by partial CCFTs. Although this metric is
a good estimation of the computational complexities, it reflects only
part of the hardware complexities. For example, buffers, multiplexers
and control units are required if we want to reuse modules to save
chip area, and their complexities need to be accounted for.  Thus, in
this section hardware implementations are used to further verify and
demonstrate the advantages of partial CCFTs.

In the literature, numerous syndrome-based RS decoder designs use the
Horner's rule \cite{BlahutECC} to implement the syndrome computation,
Chien search, and Forney's formula. Since we want to replace the
Horner's rule by partial CCFT, the syndrome computation module is
representative to illustrate the advantages of the partial
CCFT. Although the architecture and hardware design of RS decoders are
well-studied in the literature, there are few results on the RS codes
over GF$(2^{12})$ due to their extreme lengths.  Therefore, in this
section, we choose to implement in hardware the syndrome computation
block for the $(2720, 2550)$ shortened RS code in \cite{Buerner2004}
as an example, because detailed synthesis results of the syndrome
computation block are provided in \cite{Buerner2004}. Two VLSI designs
synthesized with 0.18 $\mu$m CMOS technology are provided in
\cite{Buerner2004} with different parallelization parameters. We also
implement this block with partial CCFTs, and synthesize it with a more
advanced 45 nm technology \cite{OKSU}. No hardware implementation
results is provided in \cite{Chen2009}.  Given the extreme length of
this code, since the CSE algorithm cannot be used to reduce additive
complexities of partial CFFTs, partial CCFTs have a significant
advantage against partial CFFTs, as shown in Tabs.~\ref{tab:syncomp}
and \ref{tab:ChienForney}.

\subsection{Hardware Implementations}
When we use partial CCFTs to compute the syndrome for the $(2720,
2550)$ RS code, 2720 time-domain elements and 170 frequency-domain
components are needed in the 4095-point DFT.  If we implement this
block in a fully parallel fashion, the computational complexity
in Tab.~\ref{tab:syncomp} is a good estimate of the hardware
complexity. However, the hardware complexity is too large to be
used in practice. Fortunately, the modular structure of partial CCFTs
enables us to fold the architecture. Since the CCFTs decompose the long
DFTs into several short sub-DFTs, those sub-DFTs can be used as
modules in hardware implementations. They can be reused to save the
chip area and power consumption, or pipelined and parallelized to
increase the throughput. This is a desirable property in  hardware
implementation of the RS decoders.

In our hardware implementation, we first decompose the 4095-point DFT
as $63 \times 65$ as suggested by Tab.~\ref{tab:syncomp}, i.e., first
compute 65 63-point DFTs and then compute 63 65-point DFTs. To compute these
DFTs in one clock cycle in a fully parallel way, it requires 65 63-point DFT modules and 
63 65-point DFT modules. This straightforward implementation has very high complexity. 
Instead, we carry out the partial CCFT in two steps. The
first step computes the 65 63-point DFTs in $T_1$ clock cycles, each
cycle computing at most $\lceil 65/T_1 \rceil$ 63-point DFTs; and the
second step computes the 63-point DFTs in $T_2$ clock cycles, each
cycle computing at most $\lceil 63/T_2 \rceil$ 65-point
DFTs. Therefore, we can compute the partial CCFT in $T_1+T_2$ cycles
with $\lceil 65/T_1 \rceil $ 63-point DFT modules and $\lceil 63/T2
\rceil$ 65-point DFT modules. These 63-point DFT modules and 65-point
DFT modules are implemented by CFFTs to reduce their 
complexities, and the computations involving the zero time-domain inputs
and/or unnecessary frequency-domain components are removed.

\subsection{Implementation Results and Remarks}

We provide two hardware designs with $(T_1, T_2)$ equal to $(13, 9)$
and $(5, 7)$, respectively. The synthesis results are shown in
Tab.~\ref{tab:HWcomp}, and they are compared with the two designs with
different parallelization parameters in \cite{Buerner2004}. Due to the
different process technologies used in the synthesis, the clock
rates can not be compared directly. We provide both clock rates
as well as throughputs of all implementations (the throughput is defined as the
number of vectors that can be processed in each second).
The equivalent gate count is computed by dividing the total
chip area by the area of an XOR gate in the corresponding technology,
and it can serve as a metric to compare designs in
different process technologies. 

\begin{table}[t]
  \centering
  \caption{Comparison of the VLSI implementations of the syndrome
    computation for (2720, 2550) RS code.}
  \label{tab:HWcomp}
  \begin{tabular}{|c|c|c|c|c|}
    \hline
    & \multicolumn{2}{c|}{Partial CCFT} &
    \multicolumn{2}{c|}{Honer's Rule \cite{Buerner2004}} \\
    \hline
     $(T_1, T_2)$ & $(5, 7)$ & $(13, 9)$ & &\\
    Process & 45 nm & 45 nm & 0.18 $\mu$m & 0.18 $\mu$m \\
    Clock rate & 250 MHz & 200 MHz & 112 MHz & 225 MHz \\
    Gate count & 384k & 306k & 920k & 480k \\
    Require cycles & 12 & 22 & 86 & 171 \\
    Throughput (vec/s)& 20.8M & 9.1M & 1.3M & 1.3M \\
    \hline
  \end{tabular}
\end{table}

From Tab.~\ref{tab:HWcomp}, we can see that both the gate count and
required cycles are reduced greatly compared with the designs in
\cite{Buerner2004} because a partial CCFT has a much smaller
computational complexity than Horner's rule. With partial CCFTs, we can
design an RS decoder with smaller area and larger throughput because
of reduced gate counts and required numbers of cycles, respectively.

Due to the modular structure of partial CCFTs, we can make a wide
range of trade-offs between the chip area and throughput. We can
reduce the number of the required cycles by increasing the number of
sub-DFT modules in each tier, and the chip area is therefore
increased. For example, if we reduce the required cycles from 22 to
12, the gate count increases from 306k to 384k as shown in
Tab.~\ref{tab:HWcomp}. In contrast, it is not easy for partial CFFTs
to make such trade-offs because of the irregular structure of the
post-addition network for partial CFFTs (see \cite{Chen2009}).
Moreover, since we compute the sub-DFTs by CFFTs, which are
implemented as bilinear algorithms and also have modular structure, we
can shorten the critical path and improve the clock rate by pipelining
the sub-DFT modules, i.e., inserting pipeline registers between
pre-addition network, multipliers, and post-addition network.

We remark that we focus on the decomposition $63 \times 65$ for the
4095-point DFT above. Other decompositions, even multi-tier structure
decomposition, can be considered. For example, a decomposition $7
\times 9 \times 5 \times 13$ would lead to a four-tier structure,
which leads to a smaller critical path delay since the the sub-DFTs in
each tier are smaller and they can be pipelined.

\section{Conclusion}
\label{sec:conclusion}

We extend our previous work in \cite{Wu2011a} by proposing partial
CCFT to reduce the computational complexity of syndrome based RS
decoder. Our results show that partial CCFTs have advantages in
reducing the computational complexity of the DFTs, which can be used
to implement the syndrome computation, Chien search, and Forney's
formula. The hardware implementation results show that since the
computational complexity is reduced greatly, smaller chip area and
fewer clock cycles are needed to compute the syndrome of the a received
vector. Moreover, the modular structure of partial CCFT provides a wide range of
trade-offs between the chip area and throughput, which is a favorable
property in hardware designs.

\bibliographystyle{IEEEtran}
\bibliography{IEEEabrv,FFT}

\begin{thebibliography}{10}
\providecommand{\url}[1]{#1}
\csname url@samestyle\endcsname
\providecommand{\newblock}{\relax}
\providecommand{\bibinfo}[2]{#2}
\providecommand{\BIBentrySTDinterwordspacing}{\spaceskip=0pt\relax}
\providecommand{\BIBentryALTinterwordstretchfactor}{4}
\providecommand{\BIBentryALTinterwordspacing}{\spaceskip=\fontdimen2\font plus
\BIBentryALTinterwordstretchfactor\fontdimen3\font minus
  \fontdimen4\font\relax}
\providecommand{\BIBforeignlanguage}[2]{{%
\expandafter\ifx\csname l@#1\endcsname\relax
\typeout{** WARNING: IEEEtran.bst: No hyphenation pattern has been}%
\typeout{** loaded for the language `#1'. Using the pattern for}%
\typeout{** the default language instead.}%
\else
\language=\csname l@#1\endcsname
\fi
#2}}
\providecommand{\BIBdecl}{\relax}
\BIBdecl

\bibitem{BlahutECC}
R.~E. Blahut, \emph{Theory and Practice of Error Control Codes}.\hskip 1em plus
  0.5em minus 0.4em\relax Addison-Wesley, 1984.

\bibitem{Buerner2004}
T.~Buerner, R.~Dohmen, A.~Zottmann, M.~Saeger, and A.~J. van Wijngaarden, ``On
  a high-speed {Reed-Solomon} {CODEC} architecture for 43 {Gb}/s optical
  transmission systems,'' in \emph{Proc. 24th International Conference on
  Microelectronics}, vol.~2, May 16--19, 2004, pp. 743--746.

\bibitem{Jeon2010}
S.~Jeon and {B.V.K. Vijaya Kumar}, ``Performance and complexity of 32 k-bit
  binary ldpc codes for magnetic recording channels,'' \emph{{IEEE} Trans.
  Magn.}, vol.~46, no.~6, pp. 2244--2247, 2010.

\bibitem{Han2005}
Y.~Han, W.~E. Ryan, and R.~Wesel, ``Dual-mode decoding of product codes with
  application to tape storage,'' in \emph{Proc. IEEE Global Telecommunications
  Conference}, vol.~3, Nov. 28--Dec. 2, 2005, pp. 1255--1260.

\bibitem{IDEMAWhitePaper}
\BIBentryALTinterwordspacing
``Hard disk drive long data sector white paper,'' April 20 2007. [Online].
  Available: \url{http://www.idema.org/}
\BIBentrySTDinterwordspacing

\bibitem{Costa2004}
\BIBentryALTinterwordspacing
E.~Costa, S.~V. Fedorenko, and P.~V. Trifonov, ``On computing the syndrome
  polynomial in {Reed-Solomon} decoder,'' \emph{European Transactions on
  Telecommunications}, vol.~15, no.~4, pp. 337--342, 2004. [Online]. Available:
  \url{http://dx.doi.org/10.1002/ett.982}
\BIBentrySTDinterwordspacing

\bibitem{Chen2009}
N.~Chen and Z.~Yan, ``Reduced-complexity {Reed-Solomon} decoders based on
  cyclotomic {FFTs},'' \emph{{IEEE} Signal Process. Lett.}, vol.~16, no.~4, pp.
  279--282, 2009.

\bibitem{Trifonov2003}
P.~V. Trifonov and S.~V. Fedorenko, ``A method for fast computation of the
  {Fourier} transform over a finite field,'' \emph{Probl. Inf. Transm.},
  vol.~39, no.~3, pp. 231--238, 2003.

\bibitem{Fedorenko2006}
S.~V. Fedorenko, ``A method for computation of the discrete {Fourier} transform
  over a finite fields,'' \emph{Probl. Inf. Transm.}, vol.~42, pp. 139--151,
  2006.

\bibitem{Chen2009c}
N.~Chen and Z.~Yan, ``Cyclotomic {FFT}s with reduced additive complexities
  based on a novel common subexpression elimination algorithm,'' \emph{{IEEE}
  Trans. Signal Process.}, vol.~57, no.~3, pp. 1010--1020, Mar. 2009.

\bibitem{Wu2011a}
X.~Wu, M.~Wagh, N.~Chen, Y.~Wang, and Z.~Yan, ``Composite cyclotomic fourier
  transforms with reduced complexities,'' \emph{{IEEE} Trans. Signal Process.},
  vol.~59, no.~5, pp. 2136--2145, 2011.

\bibitem{Good1958}
\BIBentryALTinterwordspacing
I.~J. Good, ``The interaction algorithm and practical {Fourier} analysis,''
  \emph{Journal of the Royal Statistical Society. Series B (Methodological)},
  vol.~20, no.~2, pp. 361--372, 1958. [Online]. Available:
  \url{http://www.jstor.org/stable/2983896}
\BIBentrySTDinterwordspacing

\bibitem{Cooley1965}
J.~W. Cooley and J.~W. Tukey, ``An algorithm for the machine calculation of
  complex {Fourier} series,'' \emph{Mathematics of Computation}, vol.~19,
  no.~90, pp. 297--301, 1965.

\bibitem{Truong2006}
T.~K. Truong, P.~D. Chen, L.~J. Wang, Y.~Chang, and I.~S. Reed, ``Fast, prime
  factor, discrete {Fourier} transform algorithms over ${GF}(2^m)$ for $8 \le m
  \le 10$,'' \emph{Inf. Sci.}, vol. 176, no.~1, pp. 1--26, 2006.

\bibitem{OKSU}
\BIBentryALTinterwordspacing
 [Online]. Available: \url{http://vcag.ecen.okstate.edu/projects/scells/}
\BIBentrySTDinterwordspacing

\end{thebibliography}

\end{document}